\documentclass[nofootinbib,showpacs,preprintnumbers,aps]{revtex4}
\usepackage {amsfonts}
\usepackage {graphicx}
\usepackage {longtable}
\usepackage{amsmath,amssymb}
\begin{document}

\title{Birefringence (rotation of polarization plane and spin dichroism) of deuterons in carbon
target}

\author{V.Baryshevsky\protect\footnote{ E-mail: bar@inp.minsk.by;}, A.
Rouba\protect\footnote{ E-mail: rouba@inp.minsk.by}}
\address{Research Institute for Nuclear Problems, Bobruiskaya Str.\@11,
220050 Minsk, Belarus}

\begin{abstract}

Birefringence  phenomenon for deuteron with energy up 20 MeV in
carbon target is considered. The estimation for spin dichroism and
for angle of rotation of polarization plane of deuterons is
presented. It is shown that magnitude of the phenomenon strongly
depends on behavior of the deuteron wave functions on small
distance between nucleon in deuteron.
\end{abstract}

\maketitle
\section{Introduction}

It was shown in \cite{Bar92}, \cite{Bar93} that  birefringence
effect  arise for deuterons (in general for all particles with
spin $S\geqslant1$) passing through unpolarized isotropic matter.
According to \cite{Bar92}, \cite{Bar93} this phenomenon is caused
by difference of spindependent forward scattering amplitude for
deuterons with spin projection $m=0$ and $m=\pm1$ on coordinate
axis parallel to deuteron wave vector ($m$ is magnetic quantum
number). As the result there is appears possibility to study real
and imaginary part of spindependent forward scattering amplitude
at experiment. The first experimental study of deuteron spin
dichroism in
 carbon target was carried out  at the electrostatic HVEC
tandem Van-de-Graaff accelerator with deuterons of up to 20~MeV (
Institut f\"{u}r Kernphysik of Universit\"at zu K\"oln)
\cite{Exp1}, \cite{Exp2}. According to experimental results
appearance of tensor polarization in the transmitted deuterons
beam was observed \cite{Exp1}, \cite{Exp2}. As a result spin
dichroism of deuteron beam passing through unpolarized carbon
target was discovered. Later in 2007  spin dichroism was measured
for 5.5 GeV/c  deuterons in carbon target on Nuclotron in Dubna
\cite{dub}.

In this paper the difference of spindependet forward scattering
amplitude for deuteron with energy up to 20 MeV on carbon target
on the base of eikonal approximation is considered. The estimation
for angle of rotation of polarization plane and for deuteron spin
dichroism is done. It is shown that magnitude of the phenomenon
strongly depends on behavior of the deuteron wave functions for
small distance between nucleon in deuteron.

\section{Eikonal approximation for deuteron spindependet forward
scattering amplitude on carbon target}

Let us discuss a possible magnitude of the deuteron birefringence
effect. According to \cite{Bar92}, \cite{Bar93} birefringence
effect depends on amplitudes of zero-angle elastic coherent
scattering of a deuteron by a nucleus $f(m=\pm1)$ and $f(m=0)$.
Let us consider theory of birefringence effect \cite{Bar92},
\cite{Bar93} briefly.

The  Hamiltonian $H$ can be written as
\begin{equation}
H=H_D(\vec{r_p},\vec{r_n})+H_D(\vec{r})+H_N(\{\xi_i\})+V_{DN}(\vec{r_p},\vec{r_n},\{\xi_i\})
\label{ham1}
\end{equation}
where $H_D$ is the deuteron Hamiltonian; $H_N$ the nuclear
Hamiltonian; $V_{DN}$ stands for the energy of deuteron-nucleus
nuclear and Coulomb interaction; $r_p(r_n)$ is the coordinate of
the proton (neutron) entering into the deuteron $\{\xi_i\}$ is a
set of coordinates of the nucleons. Introducing the deuteron
center-of-mass coordinate $\vec{R}$ and the relative distance
between the proton and neutron in the deuteron
$\vec{r}=\vec{r_p}-\vec{r_n}$, we write (\ref{ham1}) as
\begin{equation}
H=-\frac{\hbar^2}{2m_D}\Delta(\vec{R})+H_D(\vec{r})+H_N(\{\xi_i\})+V_{DN}^N(\vec{R},\vec{r},\{\xi_i\})+V_{DN}^C(\vec{R},\vec{r},\{\xi_i\})
\label{ham2}
\end{equation}
Let us consider scattering of deuterons with energy  above
deuterons binding energy $\varepsilon_d$. For deuterons with
energy 10 MeV time of nuclear deuteron-carbon interaction is
$\tau^N\simeq5\cdot10^{-22}\,s. $  whereas the characteristic
period of oscillation of nucleus in the deuteron is
$\tau\simeq2\pi\hbar/\varepsilon_d\simeq2\cdot10^{-21}\,s.$  As a
result we can apply the impulse approximation. In this
approximation we can neglect of binding energy of nucleons in
deuteron i. e. neglect of $H_D(\vec{r})$ in (\ref{ham2}). As a
result,
\begin{equation}
H=-\frac{\hbar^2}{2m_D}\Delta(\vec{R})+H_N(\{\xi_i\})+V_{DN}^N(\vec{R},\vec{r},\{\xi_i\})+V_{DN}^C(\vec{R},\vec{r},\{\xi_i\})
\label{ham3}
\end{equation}

As is seen, in the impulse approximation the problem of
determining the scattering amplitude reduces to the problem of
scattering by a nucleus of a structureless particle of the
deuteron mass. In this case the coordinate $r$ is a parameter.
Therefore, the relations obtained for the cross section and the
forward scattering amplitude should be averaged over the parameter
mentioned.  For fast deuterons the scattering amplitude can be
found in the eikonal approximation \cite{Cryz} , \cite{Hand}. The
amplitude of coherent zero-angle  scattering can be written in
this approximation as follows

\begin{eqnarray}
f(0)=\frac{k}{2\pi~i}\int \left( e^{i\chi _{D}\left( \vec{b},\vec{r}%
\right) }-1\right) d^{2}b\left| \varphi \left( \vec{r}\right)
\right| ^{2}d^{3}r \label{amp}
\end{eqnarray}

where $k$ is the deuteron wavenumber, $\vec{b}$ is the impact
parameter, $\varphi \left( \vec{r}\right)$ is deuterons wave
function. The phase shift due to the deuteron scattering by a
carbon is
\begin{equation}
\chi_{D}=\chi_{pN}+\chi_{nN}+\chi_{pN}^C=-\frac{1}{\hbar v}
\int_{-\infty }^{+\infty }V_{pN}\left( \vec{b},z',%
\vec{r}_{\perp }\right) dz'-\frac{1}{\hbar v}
\int_{-\infty }^{+\infty }V_{nN}\left( \vec{b},z',%
\vec{r}_{\perp }\right) dz'-\frac{1}{\hbar v} \int_{-\infty
}^{+\infty }V_{pN}^C\left( \vec{b},z',\vec{r}_{\perp }\right) dz',
\label{SV}
\end{equation}
where $\vec{r}_{\perp}$ is the $\vec{r}$ component perpendicular
to the momentum of incident deuteron, $v$ is the deuteron speed.

For the polarized deuteron under consideration the probability
$\left| \varphi \left( \vec{r}\right)\right| ^{2}$ is
differentiate for different spin states of deuteron. Thus for
states with magnetic quantum number $m=\pm 1$, the probability is
$\left| \varphi_{\pm 1} \left( \vec{r}\right)\right| ^{2}$,
whereas for $m=0$, it is $\left| \varphi_{0} \left(
\vec{r}\right)\right| ^{2}$. Owing to the additivity of phase
shifts, equation (\ref{amp}) can be rewritten as
\begin{eqnarray}
f\left( 0\right)&=&\frac{k}{\pi }\int \left\{ t_{pN} \left(
\vec{b}+\frac{\vec{r}_{\perp }}{2} \right) +t_{nN} \left( \vec{b}-
\frac{\vec{r}_{\perp }}{2} \right) + t_{pN}^C \left(
\vec{b}+\frac{\vec{r}_{\perp }}{2} \right)+2it_{pN} \left(
\vec{b}+\frac{\vec{r}_{\perp }}{2} \right) t_{pN}^C \left(
\vec{b}+\frac{\vec{r}_{\perp }}{2} \right)\right\}  \left| \varphi
\left( \vec{r} \right) \right|
^{2} d^{2}bd^{3}r\nonumber\\
&+&\frac{k}{\pi }\int \left\{2it_{nN} \left( \vec{b
}-\frac{\vec{r}_{\perp }}{2} \right) t_{pN}^C \left(
\vec{b}+\frac{\vec{r}_{\perp }}{2} \right) +2it_{pN} \left( \vec{b
}+\frac{\vec{r}_{\perp }}{2} \right) t_{nN}
\left(\vec{b}-\frac{\vec{r}_{\perp }}{2} \right)\right\}  \left|
\varphi \left( \vec{r} \right) \right|
^{2} d^{2}bd^{3}r\nonumber\\
&-&\frac{k}{\pi }\int 4t_{pN} \left( \vec{b }+\frac{\vec{r}_{\perp
}}{2} \right)t_{nN} \left( \vec{b }-\frac{\vec{r}_{\perp }}{2}
\right)t_{pN}^C \left( \vec{b}+\frac{\vec{r}_{\perp }}{2} \right)
\left| \varphi \left( \vec{r} \right) \right| ^{2} d^{2}bd^{3}r,
 \label{42}
\end{eqnarray} where
\[t_{nN(pN)}^{(C)}=\frac{e^{i\chi _{nN\left(pN\right)
}^{(C)}}-1}{2i}.
\]
 From
(\ref{42}) it follows
\begin{eqnarray}
f(0)&=&F_{pN}(0)+F_{nN}(0)+F_{pN}^C(0)+2iF_{ppN}^C+\frac{2ik}{\pi}
\int t_{nN}\left( \vec{b}-\frac{\vec{r}_{\perp }}{2}\right)t_{pN}^C\left( \vec{b}+%
\frac{\vec{r}_{\perp }}{2}\right)\left| \varphi \left(
\vec{r}_{\perp},z\right) \right| ^{2}d^{2}bd^{2}r_{\perp}dz\nonumber\\
&+&\frac{2ik}{\pi}
\int t_{pN}\left( \vec{b}+\frac{\vec{r}_{\perp }}{2}\right)t_{nN}\left( \vec{b}-%
\frac{\vec{r}_{\perp }}{2}\right)\left| \varphi \left(
\vec{r}_{\perp},z\right) \right| ^{2}d^{2}bd^{2}r_{\perp}dz\nonumber\\
&-&\frac{4k}{\pi} \int t_{pN} \left( \vec{b }+\frac{\vec{r}_{\perp
}}{2} \right)t_{nN} \left( \vec{b }-\frac{\vec{r}_{\perp }}{2}
\right)t_{pN}^C \left( \vec{b}+\frac{\vec{r}_{\perp }}{2}
\right)\left| \varphi \left( \vec{r}_{\perp},z\right) \right|
^{2}d^{2}bd^{2}r_{\perp}dz,\label{integral}
\end{eqnarray}
where
\[
F_{nN(pN)}^{(C)}(0)=\frac{k}{\pi} \int
t_{nN(pN)}^{(C)}(\vec{\xi})d^{2}\xi=
\frac{m_D}{m_{n(p)}}~f_{n(p)}^{(C)}(0),~~F_{ppN}^C(0)=\frac{k}{\pi}
\int t_{pN}~t_{pN}^C(\vec{\xi})d^{2}\xi
\]
and $f_{n(n)}^{(C)}(0)$ is the nuclear and the Coulomb amplitude
of the proton (neutron)-carbon zero-angle elastic coherent
scattering. The expression (\ref{integral}) can be rewritten as
\begin{eqnarray}
f(0)&=&F_{pN}(0)+F_{nN}(0)+F_{pN}^C(0)+2iF_{ppN}^C+\frac{2ik}{\pi}
\int t_{nN}\left( \vec{\eta}\right)t_{pN}^C\left(
\vec{\xi}\right)\left|
\varphi \left(\vec{\xi}-\vec{\eta},z\right) \right|^2 d^{2} \xi d^{2} {\eta}dz \\
&+&\frac{2ik}{\pi} \int t_{pN}\left(
\vec{\xi}\right)t_{nN}\left(\vec{\eta}\right)\left| \varphi \left(
\vec{\xi}-\vec{\eta},z\right) \right|^2 d^{2} \xi d^{2} {\eta}dz
-\frac{4k}{\pi} \int t_{pN} \left( \vec{\xi }\right)t_{nN} \left(
\vec{\eta}\right)t_{pN}^C \left( \vec{\xi}\right)\left| \varphi
\left(\vec{\xi}-\vec{\eta},z\right) \right| ^{2}d^{2} \xi d^{2}
{\eta}dz.\nonumber \label{27}
\end{eqnarray}

Then from (\ref{27}) {\small{
\begin{eqnarray}
Ref(0)&=&ReF_{pN}(0)+ReF_{nN}(0)+ReF_{pN}^C(0)-2ImF_{ppN}^C-\frac{2k}{\pi}
Im\int t_{nN}\left( \vec{\eta}\right)t_{pN}^C\left(
\vec{\xi}\right)\left|
\varphi \left(\vec{\xi}-\vec{\eta},z\right) \right|^2 d^{2} \xi d^{2} {\eta}dz \nonumber\\
&-&\frac{2k}{\pi} Im\int t_{pN}\left(
\vec{\xi}\right)t_{nN}\left(\vec{\eta}\right)\left| \varphi \left(
\vec{\xi}-\vec{\eta},z\right) \right|^2 d^{2} \xi d^{2} {\eta}dz
-\frac{4k}{\pi} Re\int t_{pN} \left( \vec{\xi }\right)t_{nN}
\left( \vec{\eta}\right)t_{pN}^C \left( \vec{\xi}\right)\left|
\varphi \left(\vec{\xi}-\vec{\eta},z\right) \right| ^{2}d^{2} \xi
d^{2} {\eta}dz,\nonumber \\
Imf(0)&=&ImF_{pN}(0)+ImF_{nN}(0)+ImF_{pN}^C(0)+2ReF_{ppN}^C+\frac{2k}{\pi}
Re\int t_{nN}\left( \vec{\eta}\right)t_{pN}^C\left(
\vec{\xi}\right)\left|
\varphi \left(\vec{\xi}-\vec{\eta},z\right) \right|^2 d^{2} \xi d^{2} {\eta}dz \\
&+&\frac{2k}{\pi} Re\int t_{pN}\left(
\vec{\xi}\right)t_{nN}\left(\vec{\eta}\right)\left| \varphi \left(
\vec{\xi}-\vec{\eta},z\right) \right|^2 d^{2} \xi d^{2} {\eta}dz
-\frac{4k}{\pi} Im\int t_{pN} \left( \vec{\xi }\right)t_{nN}
\left( \vec{\eta}\right)t_{pN}^C \left( \vec{\xi}\right)\left|
\varphi \left(\vec{\xi}-\vec{\eta},z\right) \right| ^{2}d^{2} \xi
d^{2} {\eta}dz.\nonumber
 \label{28}
\end{eqnarray}
}}
%%%%%%%%%%%%!!!!!!!!!!!!!!!!!!!!!!!!!!!!!!!!!!!!!!!!!!!!!!!!!!!!!!!!!!!!!!!!!!!!11

In accordance with \cite{Bar92}-\cite{Exp2} rotation of
polarization plane is determined by the difference of the
amplitudes $Ref(m=\pm1)$ and $Ref(m=0)$ and spin dichroism is
determined by the difference of the amplitudes $Imf(m=\pm1)$ and
$Imf(m=0)$. From (\ref{28}) it follows that
\begin{eqnarray}
Re(d_1)&=&-\frac{2k}{\pi} Im\int t_{nN}\left(
\vec{\eta}\right)t_{pN}^C\left( \vec{\xi}\right)\left[ \left|
\varphi_{\pm1} \left(\vec{\xi}-\vec{\eta},z\right) \right|^2 -
\left| \varphi_{0} \left(\vec{\xi}-\vec{\eta},z\right)
\right|^2\right] d^{2} \xi d^{2}{\eta}dz\nonumber\\
&-&\frac{2k}{\pi} Im\int t_{pN}\left(
\vec{\xi}\right)t_{nN}\left(\vec{\eta}\right)\left[ \left|
\varphi_{\pm1} \left(\vec{\xi}-\vec{\eta},z\right) \right|^2 -
\left| \varphi_{0} \left(\vec{\xi}-\vec{\eta},z\right)
\right|^2\right] d^{2} \xi d^{2}{\eta}dz \nonumber\\
&-&\frac{4k}{\pi} Re\int t_{pN} \left( \vec{\xi }\right)t_{nN}
\left( \vec{\eta}\right)t_{pN}^C \left( \vec{\xi}\right)\left[
\left| \varphi_{\pm1} \left(\vec{\xi}-\vec{\eta},z\right)
\right|^2 - \left| \varphi_{0} \left(\vec{\xi}-\vec{\eta},z\right)
\right|^2\right] d^{2} \xi d^{2}{\eta}dz,\nonumber \\
Im(d_1)&=&\frac{2k}{\pi} Re\int t_{nN}\left(
\vec{\eta}\right)t_{pN}^C\left( \vec{\xi}\right)\left[ \left|
\varphi_{\pm1} \left(\vec{\xi}-\vec{\eta},z\right) \right|^2 -
\left| \varphi_{0} \left(\vec{\xi}-\vec{\eta},z\right)
\right|^2\right] d^{2} \xi d^{2}{\eta}dz\nonumber\\
&+& \frac{2k}{\pi} Re\int t_{pN}\left(
\vec{\xi}\right)t_{nN}\left(\vec{\eta}\right)\left[ \left|
\varphi_{\pm1} \left(\vec{\xi}-\vec{\eta},z\right) \right|^2 -
\left| \varphi_{0} \left(\vec{\xi}-\vec{\eta},z\right)
\right|^2\right] d^{2} \xi d^{2}{\eta}dz \nonumber\\
&-&\frac{4k}{\pi} Im\int t_{pN} \left( \vec{\xi }\right)t_{nN}
\left( \vec{\eta}\right)t_{pN}^C \left( \vec{\xi}\right)\left[
\left| \varphi_{\pm1} \left(\vec{\xi}-\vec{\eta},z\right)
\right|^2 - \left| \varphi_{0} \left(\vec{\xi}-\vec{\eta},z\right)
\right|^2\right] d^{2} \xi d^{2}{\eta}dz, \label{d1}
\end{eqnarray}
where $d_1$ is  the difference of spindependent forward scattering
 amplitudes.

At scattering of deuterons on light nucleus the characteristic
radius of the deuteron is large comparing with the radius of
nucleus. For this reason for estimation of effects, when
integrating, we can suppose that the functions $t_{pN}$ and
$t_{nN}$ act on $\varphi$ as $\delta$-function. Then
\begin{eqnarray}
Re(d_1)&=&-2Im \left\{F_{nN}(0)\int t_{pN}^C\left(
\vec{\xi}\right)\left[ \left| \varphi_{\pm1}
\left(\vec{\xi},z\right) \right|^2 - \left| \varphi_{0}
\left(\vec{\xi},z\right) \right|^2\right] d^{2} \xi
dz\right\}\nonumber\\
&-& \frac{2\pi}{k} Im \left\{ F_{pN}(0)F_{nN}(0)\int \left[ \left|
\varphi_{\pm1} \left(0,z\right) \right|^2 - \left| \varphi_{0}
\left(0,z\right) \right|^2\right]
dz\right\} \nonumber\\
&-&\frac{4\pi}{k} Re \left\{ F_{ppN}^C(0)F_{nN}(0)\int \left[
\left| \varphi_{\pm1} \left(0,z\right) \right|^2 - \left|
\varphi_{0} \left(0,z\right) \right|^2\right]
dz\right\},\nonumber \\
Im(d_1)&=&2Re \left\{F_{nN}(0)\int t_{pN}^C\left(
\vec{\xi}\right)\left[ \left| \varphi_{\pm1}
\left(\vec{\xi},z\right) \right|^2 - \left| \varphi_{0}
\left(\vec{\xi},z\right) \right|^2\right] d^{2} \xi
dz\right\}\nonumber\\
&+& \frac{2\pi}{k} Re \left\{ F_{pN}(0)F_{nN}(0)\int \left[ \left|
\varphi_{\pm1} \left(0,z\right) \right|^2 - \left| \varphi_{0}
\left(0,z\right) \right|^2\right]
dz\right\} \nonumber\\
&-&\frac{4\pi}{k} Im \left\{ F_{ppN}^C(0)F_{nN}(0)\int \left[
\left| \varphi_{\pm1} \left(0,z\right) \right|^2 - \left|
\varphi_{0} \left(0,z\right) \right|^2\right]
dz\right\}.\label{d1_1}
\end{eqnarray}

The magnitude of the birefringence effect is determined by
difference $  \left| \varphi_{\pm1}
\left(\vec{\xi}-\vec{\eta},z\right) \right|^2 - \left| \varphi_{0}
\left(\vec{\xi}-\vec{\eta},z\right) \right|^2 $ i.e. by the
difference of distributions of nucleon density in the deuteron for
different deuteron spin orientations. The structure of the
wavefunction $\varphi_{m}$ is well known:
\begin{equation}
\varphi_{m}=\frac{1}{\sqrt{4 \pi}} \left\{
\frac{u(r)}{r}+\frac{1}{\sqrt 8}\frac{W(r)}{r}\hat{S}_{12}
\right\} \chi_{m} \label{phi_m}
\end{equation}
where $u(r)$ is the deuteron radial wave function corresponding to
the S-wave; $W(r)$ is the radial function corresponding to the
D-wave; the operator $\hat{S}_{12}=6(\hat{\vec{S}}
\vec{n}_{r})^2-2\hat{\vec{S}}^2$; $\vec{n}_{r}=\frac{\vec{r}}{r}$;
$\hat{\vec{S}}=\frac{1}{2}(\vec{\sigma}_1+\vec{\sigma}_2)$ and
$\vec{\sigma}_{1(2)}$ ate the Pauli spin matrices describing
proton(neutron) spin.

Using (\ref{phi_m}) we obtain
\begin{eqnarray}
\left| \varphi_{\pm1} \right|^2 - \left| \varphi_{0}
 \right|^2=-\frac{3}{4 \pi} \left\{
\frac{1}{\sqrt{2}}\frac{u(r)W(r)}{r^2}-\frac{1}{4}\frac{W(r)^2}{r^2}
\right\}\left(n_{rx}^2+n_{ry}^2-2n_{rz}^2\right),\nonumber\\
\left| \varphi_{\pm1} \left(\vec{\xi},z\right) \right|^2 - \left|
\varphi_{0} \left(\vec{\xi},z\right) \right|^2=-\frac{3}{4 \pi}
\left\{
\frac{1}{\sqrt{2}}\frac{u(r)W(r)}{r^2}-\frac{1}{4}\frac{W(r)^2}{r^2}
\right\}\frac{\xi^2-2z^2}{r^2},\nonumber\\
\int \left[ \left| \varphi_{\pm1} \left(0,z\right) \right|^2 -
\left| \varphi_{0} \left(0,z\right) \right|^2\right]
dz=\frac{3}{\pi}\int_{0}^{\infty}\left\{
\frac{1}{\sqrt{2}}\frac{u(r)W(r)}{r^2}-\frac{1}{4}\frac{W(r)^2}{r^2}
\right\}dr=\frac{3}{\pi}~G,
 \label{dfi}
\end{eqnarray}
where $r^2=\xi^2+z^2$.

Substituting equation (\ref{dfi}) into (\ref{d1_1}) yields
\begin{eqnarray}
Re(d_1)&=&\frac{3}{2\pi} Im \left\{F_{nN}(0)\int t_{pN}^C\left(
\vec{\xi}\right)\left\{
\frac{1}{\sqrt{2}}\frac{u(r)W(r)}{r^2}-\frac{1}{4}\frac{W(r)^2}{r^2}
\right\}\frac{\xi^2-2z^2}{r^2} d^{2} \xi
dz\right\}\nonumber\\
&-& \frac{6}{k} Im \left\{ F_{pN}(0)F_{nN}(0)\right\}G
-\frac{12}{k} Re \left\{ F_{ppN}^C(0)F_{nN}(0)\right\}G,\nonumber \\
Im(d_1)&=&-\frac{3}{2\pi} Re \left\{F_{nN}(0)\int t_{pN}^C\left(
\vec{\xi}\right)\left\{
\frac{1}{\sqrt{2}}\frac{u(r)W(r)}{r^2}-\frac{1}{4}\frac{W(r)^2}{r^2}
\right\}\frac{\xi^2-2z^2}{r^2} d^{2} \xi
dz\right\}\nonumber\\
&+& \frac{6}{k} Re \left\{ F_{pN}(0)F_{nN}(0)\right\}G
-\frac{12}{k} Im \left\{ F_{ppN}^C(0)F_{nN}(0)\right\}G.\label{13}
\end{eqnarray}

Now we can evaluate the deuteron spin dichroism and angle of
rotation of polarization plane. Let unpolarized deuterons beam
pass through carbon target. According to \cite{Bar92}-\cite{Exp2}
spin dichroism A and tensor polarization can be written as
\begin{equation}
p_{zz} \approx - \frac{{4}}{{3}}\textrm{A}, \quad p_{xx} = p_{yy}
\approx \frac{{2}}{{3}}\textrm{A},
\end{equation}
where
\begin{equation}
\textrm{A} = \frac{{I_{0} - I_{ \pm} } }{{I_{0} + I_{ \pm} }
}=\frac{{N_a z}}{{2M_r}}\left(\sigma_{\pm1}-\sigma_0\right) =
\frac{{2\pi N_a z}}{{kM_r}}Im(d_1)\, \label{dich}
\end{equation}
${I}_{0}$ is the intensity of the deuteron beam after the target
if the deuteron beam before the target is in the spin state
${m}=0$ and, similarly, ${I}_{\pm} $ is the intensity of the
deuteron beam after the target if the deuteron beam before the
target is in the spin state $m=\pm 1$,  $z$ is thickness of target
in $g/cm^2$, $N_a$ is Avogadro number, $M_r$ is molar mass for
targets matter, $\sigma_{\pm1}$ and $\sigma_{\pm1}$ are the
deuteron total cross-section of scattering for spin state $m=\pm
1$ and $m=0$ respectively.

According to \cite{Bar92}-\cite{Exp2} angle of rotation of
polarization plane is

\begin{equation}
{\vartheta}=\frac{{2\pi N_a z}}{{kM_r}}Re(d_1).\label{rot}
\end{equation}

 For estimation of nucleon-carbon strong
interaction in (\ref{SV}) lets consider optical Woods-Saxon
potential for 5.25 MeV nucleons
$V_{nN}(r)=V_{pN}(r)=\frac{-52.5-0.9i}{1+exp\left(2\left(r-3.045\right)\right)}$.
Total cross-section for n-C scattering at 5.25 MeV calculated by
optical potential and eikonal approximation is about 1.2 barn that
agree with experimental data. For Coulomb p-C interaction in
(\ref{SV}) we consider Coulomb screening potential. For
calculation of parameter $G$   the deuterons wave functions from
\cite{Mach} was applied. Obtained value $G$ is about $0.05$.

\begin{figure}[!h]
\includegraphics[width=17cm,keepaspectratio]{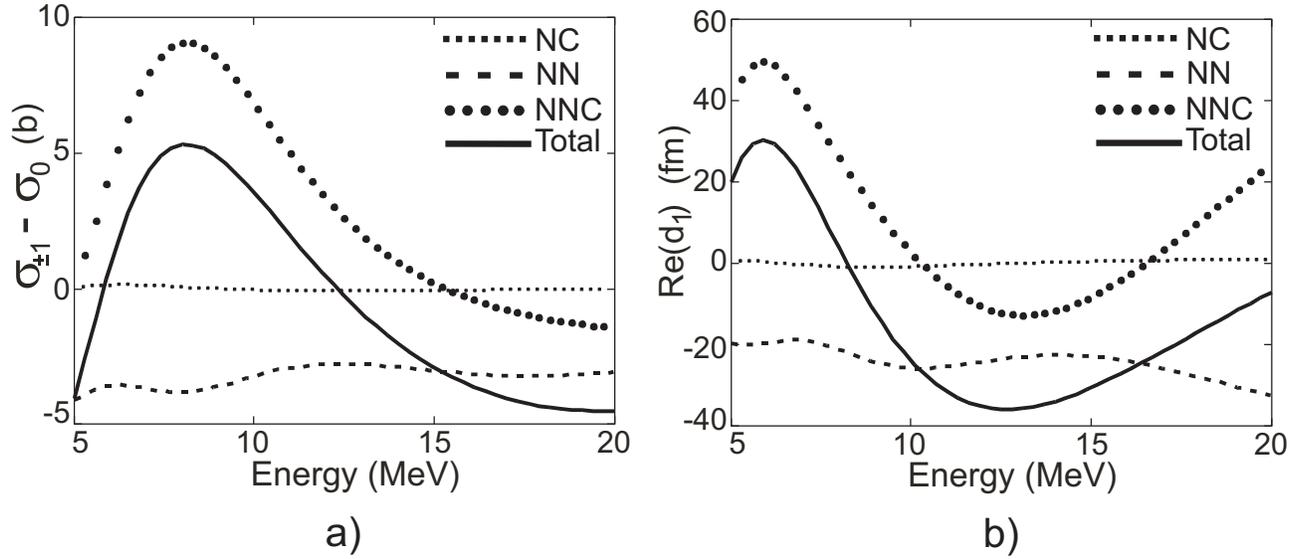}
\caption{Dependencies on deuteron energy of contributions of
items NC, NN, NNC and their sum to a) $\sigma_{\pm1}-\sigma_0$ and
b) $Re(d_1)$.} \label{fig1}
\end{figure}

In (\ref{13}) the first  items for $Re(d_1)$ and $Im(d_1)$ are
describe contribution of interference of nuclear n-C and Coulomb
p-C interactions (lets denote that as NC), the second items are
describe contribution of interference of nuclear p-C and n-C
interactions (NN) and the third items are describe contribution of
interference of nuclear p-C, n-C and Coulomb p-C interactions
(NNC). Dependencies on energy of contributions of every items to
$\sigma_{\pm1}-\sigma_0$ and $Re(d_1)$  are shown on
fig.\ref{fig1}.

\begin{figure}[!h]
\includegraphics[width=7cm,keepaspectratio]{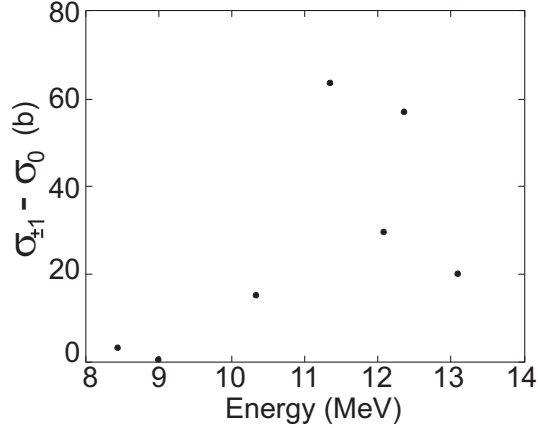}
\caption{Dependence of averaged effective difference of total
cross-section $\sigma_{\pm1}-\sigma_0$ on averaged deuteron energy
inside carbon targets obtained in experiments \cite{Exp1},
\cite{Exp2}.}
\label{fig4}
\end{figure}

  So according
(\ref{dich}) for carbon target with $z=0.1\,g/cm^2$  and for
energy conditions  of experiment (6-13 MeV) \cite{Exp1},
\cite{Exp2} dichroism is about 0.01. On the fig.\ref{fig4} is
shown dependence of averaged effective difference of total
cross-section $\sigma_{\pm1}-\sigma_0$ on averaged deuteron energy
inside carbon targets obtained in experiments \cite{Exp1},
\cite{Exp2}. We would like to make a note that since
$\sigma_{0}\neq\sigma_{\pm1}$, measuring of spin dichroism
possible in spin-filtering experiment in COSY (investigation of
different decreasing of intensity of transmitted proton beam after
passing of deuteron target with longitudinal and transversal
polarization relatively proton wave vector).

For same carbon target but for a little higher energy angle of
rotation of polarization plane is about -0.5$^\circ$. So that
value of rotation can be measured experimentally on the
installation described in \cite{Exp1}, \cite{Exp2}.

\begin{figure}[!h]
\includegraphics[width=17cm,keepaspectratio]{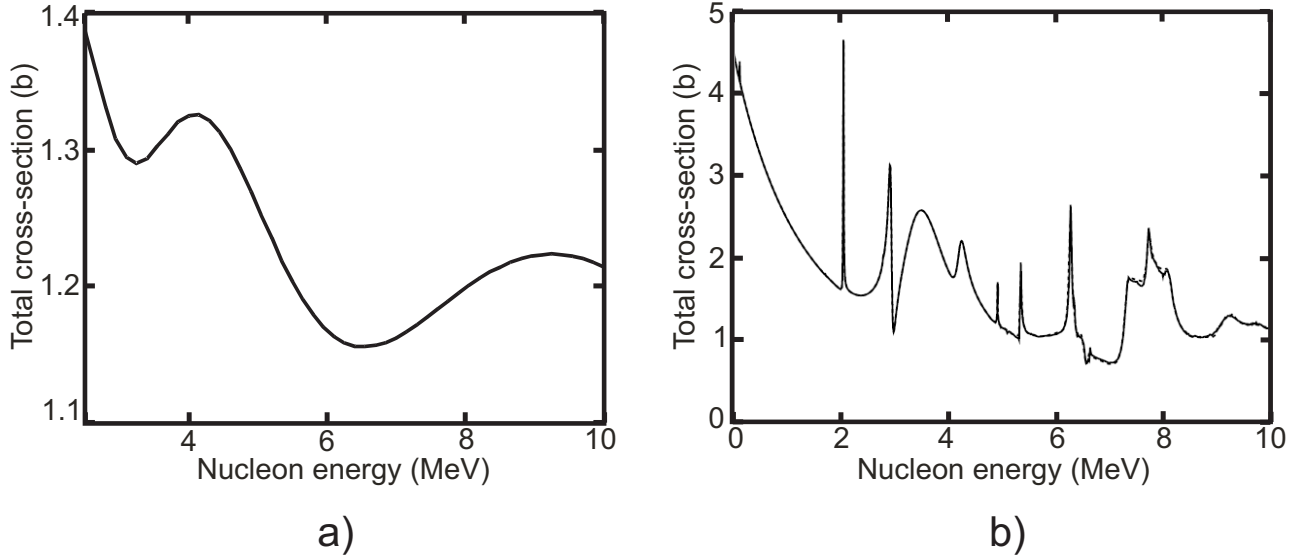}
\caption{Dependencies on nucleon energy of total nucleon-carbon
cross-section calculated by  a) eikonal approximation and b) from
experimental data.} \label{fig2}
\end{figure}

There are some reasons that can essentially increase birefringence
effect. First of all, it is interaction of nucleon with carbon. On
the fig.\ref{fig2} is shown the estimated total cross-section,
calculated by simple Woods-Saxon potential and eikonal
approximation in comparison with  experimental total
cross-section. Interaction of nucleon with carbon has a lot of
resonances in energy region of carried out experiment. So
experimental cross-section for some energy interval in 2-2.5 times
more than estimated that can result in increasing of effects up to
4-6.25 times for that energy interval. At the second, parameter G
is very sensitive to deuterons wave functions at small distances.
At the third, the increasing of weight of D-state (in \cite{Mach}
it is 4.85\%) is increase birefringence effects.

According to fig.\ref{fig1} Coulomb scattering play very important
role in birefringence value and behavior. Position of peak, caused
by Coulomb interaction is sensitive to Coulomb potential so it can
be shifted for realistic interaction. Fig.\ref{fig1} and
fig.\ref{fig2} give qualitative explanation of experimental
results on fig.\ref{fig4} \cite{Exp1}, \cite{Exp2}: sign of
dichroism, strong dependence on energy, non-monotone and
non-linear dependence of dichroism on target thickness.

Let's consider now another model of d-C interaction. At deuterons
energy 10.5 MeV the characteristic time of Coulomb deuteron-carbon
interaction caused by radius of Coulomb screening is
$\tau^C\simeq2\cdot10^{-18}\,s.$  i. e. much more then
characteristic period of nucleons oscillation in deuteron $\tau$
($R^C\simeq3\cdot10^{-9}\,cm \gg R^N\simeq3\cdot10^{-13}\,cm$
where $R^C$ and $R^N$ are radiuses of Coulomb screening and
nucleus of carbon respectively). So in this case we can't to
neglect by deuterons binding energy. Due to high frequency of
oscillation of nucleons in carbon and in deuteron for
deuteron-carbon Coulomb interaction we can use approximation for
that in (\ref{ham2}) Coulomb interaction is averaged by deuterons
ground state wave functions. I this case Coulomb potential of p-C
interaction will be shifted from the proton to the deuteron center
of mass. Then (\ref{42}) can be  written as
\begin{eqnarray}
f\left( 0\right)&=&\frac{k}{\pi }\int \left\{ t_{pN} \left(
\vec{b}+\frac{\vec{r}_{\perp }}{2} \right) +t_{nN} \left( \vec{b}-
\frac{\vec{r}_{\perp }}{2} \right) + t_{pN}^C \left(
b\right)+2it_{pN} \left( \vec{b}+\frac{\vec{r}_{\perp }}{2}
\right) t_{pN}^C \left( b\right)\right\} \left| \varphi \left(
\vec{r} \right) \right|
^{2} d^{2}bd^{3}r\nonumber\\
&+&\frac{k}{\pi }\int \left\{2it_{nN} \left( \vec{b
}-\frac{\vec{r}_{\perp }}{2} \right) t_{pN}^C \left( b\right)
+2it_{pN} \left( \vec{b }+\frac{\vec{r}_{\perp }}{2} \right)
t_{nN} \left(\vec{b}-\frac{\vec{r}_{\perp }}{2} \right)\right\}
\left| \varphi \left( \vec{r} \right) \right|
^{2} d^{2}bd^{3}r\nonumber\\
&-&\frac{k}{\pi }\int 4t_{pN} \left( \vec{b }+\frac{\vec{r}_{\perp
}}{2} \right)t_{nN} \left( \vec{b }-\frac{\vec{r}_{\perp }}{2}
\right)t_{pN}^C \left(b\right) \left| \varphi \left( \vec{r}
\right) \right| ^{2} d^{2}bd^{3}r.
 \label{42a}
\end{eqnarray}

From (\ref{42a}) it follows
\begin{eqnarray}
f(0)&=&F_{pN}(0)+F_{nN}(0)+F_{pN}^C(0)+\frac{2ik}{\pi} \int\left\{
t_{pN}\left( \vec{b}+\frac{\vec{r}_{\perp
}}{2}\right)t_{pN}^C\left(b\right)+t_{nN}\left(
\vec{b}-\frac{\vec{r}_{\perp
}}{2}\right)t_{pN}^C\left(b\right)\right\}\left| \varphi \left(
\vec{r}_{\perp},z\right) \right| ^{2}d^{2}bd^{2}r_{\perp}dz\nonumber\\
&+&\frac{2ik}{\pi}
\int t_{pN}\left( \vec{b}+\frac{\vec{r}_{\perp }}{2}\right)t_{nN}\left( \vec{b}-%
\frac{\vec{r}_{\perp }}{2}\right)\left| \varphi \left(
\vec{r}_{\perp},z\right) \right| ^{2}d^{2}bd^{2}r_{\perp}dz\nonumber\\
&-&\frac{4k}{\pi} \int t_{pN} \left( \vec{b }+\frac{\vec{r}_{\perp
}}{2} \right)t_{nN} \left( \vec{b }-\frac{\vec{r}_{\perp }}{2}
\right)t_{pN}^C \left(b\right)\left| \varphi \left(
\vec{r}_{\perp},z\right) \right|
^{2}d^{2}bd^{2}r_{\perp}dz.\label{integral_a}
\end{eqnarray}
The expression (\ref{integral_a}) can be rewritten as {\small{
\begin{eqnarray}
f(0)&=&F_{pN}(0)+F_{nN}(0)+F_{pN}^C(0)+\frac{2ik}{\pi}
\int\left\{t_{pN}\left( \vec{\xi}\right)t_{pN}^C\left(
\frac{\vec{\xi}+\vec{\eta}}{2}\right)+ t_{nN}\left(
\vec{\eta}\right)t_{pN}^C\left(
\frac{\vec{\xi}+\vec{\eta}}{2}\right)\right\}\left|
\varphi \left(\vec{\xi}-\vec{\eta},z\right) \right|^2 d^{2} \xi d^{2} {\eta}dz \nonumber\\
&+&\frac{2ik}{\pi} \int t_{pN}\left(
\vec{\xi}\right)t_{nN}\left(\vec{\eta}\right)\left| \varphi \left(
\vec{\xi}-\vec{\eta},z\right) \right|^2 d^{2} \xi d^{2} {\eta}dz
-\frac{4k}{\pi} \int t_{pN} \left( \vec{\xi }\right)t_{nN} \left(
\vec{\eta}\right)t_{pN}^C \left(
\frac{\vec{\xi}+\vec{\eta}}{2}\right)\left| \varphi
\left(\vec{\xi}-\vec{\eta},z\right) \right| ^{2}d^{2} \xi d^{2}
{\eta}dz. \label{27_a}
\end{eqnarray}
}}
 Then from (\ref{27_a})
{\small{
\begin{eqnarray}
Ref(0)&=&ReF_{pN}(0)+ReF_{nN}(0)+ReF_{pN}^C(0)\nonumber\\
&-&\frac{2k}{\pi} Im\int\left\{t_{pN}\left(
\vec{\xi}\right)t_{pN}^C\left(
\frac{\vec{\xi}+\vec{\eta}}{2}\right)+ t_{nN}\left(
\vec{\eta}\right)t_{pN}^C\left(
\frac{\vec{\xi}+\vec{\eta}}{2}\right)\right\}\left|
\varphi \left(\vec{\xi}-\vec{\eta},z\right) \right|^2 d^{2} \xi d^{2} {\eta}dz \nonumber\\
&-&\frac{2k}{\pi} Im\int t_{pN}\left(
\vec{\xi}\right)t_{nN}\left(\vec{\eta}\right)\left| \varphi \left(
\vec{\xi}-\vec{\eta},z\right) \right|^2 d^{2} \xi d^{2} {\eta}dz
-\frac{4k}{\pi} Re\int t_{pN} \left( \vec{\xi }\right)t_{nN}
\left( \vec{\eta}\right)t_{pN}^C \left(
\frac{\vec{\xi}+\vec{\eta}}{2}\right)\left| \varphi
\left(\vec{\xi}-\vec{\eta},z\right) \right| ^{2}d^{2} \xi
d^{2} {\eta}dz,\nonumber \\
Imf(0)&=&ImF_{pN}(0)+ImF_{nN}(0)+ImF_{pN}^C(0)\nonumber\\
&+&\frac{2k}{\pi} Re\int\left\{t_{pN}\left(
\vec{\xi}\right)t_{pN}^C\left(
\frac{\vec{\xi}+\vec{\eta}}{2}\right)+ t_{nN}\left(
\vec{\eta}\right)t_{pN}^C\left(
\frac{\vec{\xi}+\vec{\eta}}{2}\right)\right\}\left|
\varphi \left(\vec{\xi}-\vec{\eta},z\right) \right|^2 d^{2} \xi d^{2} {\eta}dz \\
&+&\frac{2k}{\pi} Re\int t_{pN}\left(
\vec{\xi}\right)t_{nN}\left(\vec{\eta}\right)\left| \varphi \left(
\vec{\xi}-\vec{\eta},z\right) \right|^2 d^{2} \xi d^{2} {\eta}dz
-\frac{4k}{\pi} Im\int t_{pN} \left( \vec{\xi }\right)t_{nN}
\left( \vec{\eta}\right)t_{pN}^C \left(
\frac{\vec{\xi}+\vec{\eta}}{2}\right)\left| \varphi
\left(\vec{\xi}-\vec{\eta},z\right) \right| ^{2}d^{2} \xi d^{2}
{\eta}dz.\nonumber
 \label{28_a}
\end{eqnarray}
}} From (\ref{28_a}) for spin-dependent part of forward scattering
amplitude follows that
\begin{eqnarray}
Re(d_1)&=&-\frac{2k}{\pi} Im\int\left\{t_{pN}\left(
\vec{\xi}\right)t_{pN}^C\left(
\frac{\vec{\xi}+\vec{\eta}}{2}\right) +t_{nN}\left(
\vec{\eta}\right)t_{pN}^C\left(
\frac{\vec{\xi}+\vec{\eta}}{2}\right)\right\}\left[ \left|
\varphi_{\pm1} \left(\vec{\xi}-\vec{\eta},z\right) \right|^2 -
\left| \varphi_{0} \left(\vec{\xi}-\vec{\eta},z\right)
\right|^2\right] d^{2} \xi d^{2}{\eta}dz\nonumber\\
&-&\frac{2k}{\pi} Im\int t_{pN}\left(
\vec{\xi}\right)t_{nN}\left(\vec{\eta}\right)\left[ \left|
\varphi_{\pm1} \left(\vec{\xi}-\vec{\eta},z\right) \right|^2 -
\left| \varphi_{0} \left(\vec{\xi}-\vec{\eta},z\right)
\right|^2\right] d^{2} \xi d^{2}{\eta}dz \nonumber\\
&-&\frac{4k}{\pi} Re\int t_{pN} \left( \vec{\xi }\right)t_{nN}
\left( \vec{\eta}\right)t_{pN}^C \left(
\frac{\vec{\xi}+\vec{\eta}}{2}\right)\left[ \left| \varphi_{\pm1}
\left(\vec{\xi}-\vec{\eta},z\right) \right|^2 - \left| \varphi_{0}
\left(\vec{\xi}-\vec{\eta},z\right)
\right|^2\right] d^{2} \xi d^{2}{\eta}dz,\nonumber \\
Im(d_1)&=&\frac{2k}{\pi} Re\int\left\{t_{pN}\left(
\vec{\xi}\right)t_{pN}^C\left(
\frac{\vec{\xi}+\vec{\eta}}{2}\right)+t_{nN}\left(
\vec{\eta}\right)t_{pN}^C\left(
\frac{\vec{\xi}+\vec{\eta}}{2}\right)\right\}\left[ \left|
\varphi_{\pm1} \left(\vec{\xi}-\vec{\eta},z\right) \right|^2 -
\left| \varphi_{0} \left(\vec{\xi}-\vec{\eta},z\right)
\right|^2\right] d^{2} \xi d^{2}{\eta}dz\nonumber\\
&+& \frac{2k}{\pi} Re\int t_{pN}\left(
\vec{\xi}\right)t_{nN}\left(\vec{\eta}\right)\left[ \left|
\varphi_{\pm1} \left(\vec{\xi}-\vec{\eta},z\right) \right|^2 -
\left| \varphi_{0} \left(\vec{\xi}-\vec{\eta},z\right)
\right|^2\right] d^{2} \xi d^{2}{\eta}dz \nonumber\\
&-&\frac{4k}{\pi} Im\int t_{pN} \left( \vec{\xi }\right)t_{nN}
\left( \vec{\eta}\right)t_{pN}^C \left(
\frac{\vec{\xi}+\vec{\eta}}{2}\right)\left[ \left| \varphi_{\pm1}
\left(\vec{\xi}-\vec{\eta},z\right) \right|^2 - \left| \varphi_{0}
\left(\vec{\xi}-\vec{\eta},z\right) \right|^2\right] d^{2} \xi
d^{2}{\eta}dz. \label{d1_a}
\end{eqnarray}

Taking into account (\ref{dfi}) and the same assumption as in
(\ref{d1_1}) we are obtain

\begin{eqnarray}
Re(d_1)&=&\frac{3}{\pi} Im \left\{F_{nN}(0)\int
t_{pN}^C\left(\frac{ \vec{\xi}}{2}\right)\left\{
\frac{1}{\sqrt{2}}\frac{u(r)W(r)}{r^2}-\frac{1}{4}\frac{W(r)^2}{r^2}
\right\}\frac{\xi^2-2z^2}{r^2} d^{2} \xi
dz\right\}\nonumber\\
&-& \frac{6}{k} Im \left\{ F_{pN}(0)F_{nN}(0)\right\}G
-\frac{12k}{\pi^2}G~ Re\int t_{pN} \left( \vec{\xi }\right)t_{nN}
\left( \vec{\eta}\right)t_{pN}^C \left(
\frac{\vec{\xi}+\vec{\eta}}{2}\right)d^{2} \xi d^{2}{\eta}dz,\nonumber \\
Im(d_1)&=&-\frac{3}{\pi} Re \left\{F_{nN}(0)\int t_{pN}^C\left(
\frac{ \vec{\xi}}{2}\right)\left\{
\frac{1}{\sqrt{2}}\frac{u(r)W(r)}{r^2}-\frac{1}{4}\frac{W(r)^2}{r^2}
\right\}\frac{\xi^2-2z^2}{r^2} d^{2} \xi
dz\right\}\nonumber\\
&+& \frac{6}{k} Re \left\{ F_{pN}(0)F_{nN}(0)\right\}G
-\frac{12k}{\pi^2}G~ Im\int t_{pN} \left( \vec{\xi }\right)t_{nN}
\left( \vec{\eta}\right)t_{pN}^C \left(
\frac{\vec{\xi}+\vec{\eta}}{2}\right)d^{2} \xi
d^{2}{\eta}dz.\label{13_a}
\end{eqnarray}

\begin{figure}[!h]
\includegraphics[width=17cm,keepaspectratio]{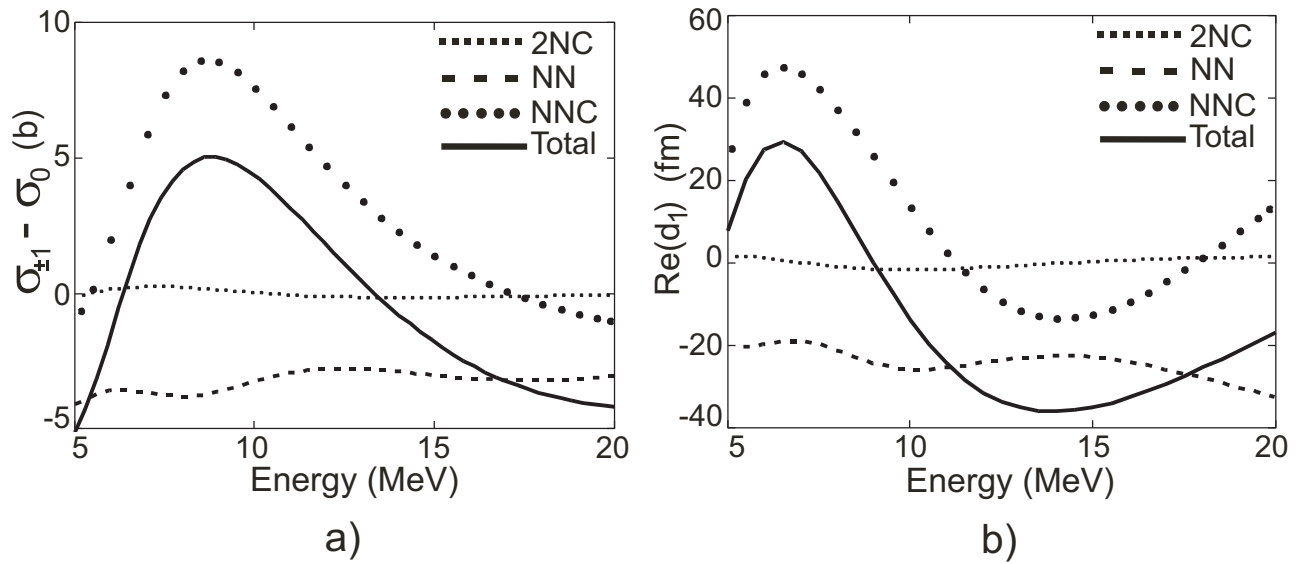}
\caption{Dependencies on deuteron energy of contributions of items
2NC, NN, NNC and their sum to a) $\sigma_{\pm1}-\sigma_0$ and b)
$Re(d_1)$.} \label{fig3}
\end{figure}

In (\ref{13_a}) the first  items for $Re(d_1)$ and $Im(d_1)$ are
describe  sum of contributions of interference of nuclear n-C
interaction with averaged Coulomb p-C interaction and nuclear p-C
interaction with averaged Coulomb p-C interaction  (lets denote
that as 2NC), the second items are describe contribution of
interference of nuclear p-C and n-C interactions (NN) and the
third items are describe contribution of interference of nuclear
p-C, n-C and averaged Coulomb p-C interactions (NNC). Dependencies
on energy of contributions of every items to
$\sigma_{\pm1}-\sigma_0$ and $Re(d_1)$ are shown on
fig.\ref{fig3}.

So, according to fig.\ref{fig1} and fig.\ref{fig3} the second
model for estimation of deuteron-carbon spin-dependent forward
scattering amplitude gives quantitatively almost the same result
as the first model but the peaks caused by Coulomb interaction is
shifted almost on 1 MeV.
\section{Summery}

Obtained results show that estimation of spin dichroism is
coincide with experimental results for some targets. More over it
necessary to note that birefringence (through the parameter G) is
very sensitive to deuterons wave functions especially at small
distances.  Here was used deuterons wave functions from
\cite{Mach}, based on CD-Bonn potential of nucleon-nucleon
interaction. According to \cite{Mach} these functions have
discrepancies with wave functions based on the another models for
$r<2\,fm.$ Especially a big problem arise with description of
deuterons wave functions for $r<0.5\,fm.$ So birefringence can be
used as additional source of information about deuterons wave
functions at small distances.

\end{document}